\documentclass[runningheads,colorBG]{svmult}
\usepackage{makeidx}   
\usepackage{graphicx}  
\usepackage{subeqnar}  
\usepackage{multicol}  
\usepackage{cropmark} 
\usepackage{physprbb}  
\newcommand{\newc}{\newcommand} 
\newc{\beq}{\begin{equation}} 
\newc{\eeq}{\end{equation}} 
\newc{\barr}{\begin{eqnarray}} 
\newc{\earr}{\end{eqnarray}} 

\usepackage{axodraw} 
\def\e6{$E(6)$}
\def\10{$SO(10)$}
\def\21{$SU(2) \otimes U(1) $}

\def\422{$SU(4) \otimes SU(2) \otimes SU(2)$}
\def\321{$SU(3) \otimes SU(2) \otimes U(1)$}
\def\ne{\hbox{$\nu_e$ }}
\def\nm{\hbox{$\nu_\mu$ }}
\def\bne{\hbox{$\bar\nu_e$ }}
\def\bnm{\hbox{$\bar\nu_\mu$ }}
\def\nt{\hbox{$\nu_\tau$ }}
\def\ns{\hbox{$\nu_{s}$ }}

\def\ed{\end{document}}
\def \nbb {$\beta\beta_{0\nu}$ }
\def\Eb{E_{\rm b}}
\def\Ee{\langle E_{\bar\nu_{\rm e}}\rangle}

\begin{document}
%
\title*{\Red  {Neutrinos: Summarizing the State-of-the-Art }}
\toctitle{ Neutrinos}
%
%
\titlerunning{ Neutrinos}
%
\author{J. W. F. Valle}
\authorrunning{J. W. F. Valle}
%
%
\institute{Instituto de F\'{\i}sica Corpuscular -- C.S.I.C., 
Universitat de Val{\`e}ncia \\
Edificio Institutos, Aptdo.\ 22085, E--46071 Val{\`e}ncia, Spain\\
E-mail:valle@ific.uv.es}
\maketitle              

\begin{abstract}
    
  I review oscillation solutions to the neutrino anomalies and discuss
  how to account for the required pattern of neutrino masses and
  mixings from first principles. Unification and low-energy bottom-up
  approaches are discussed, the latter open up the possibility of
  testing neutrino mixing at high energy colliders, such as the LHC.
  Large \ne mixing is consistent with Supernova (SN) astrophysics and
  may serve to probe galactic SN parameters at Cherenkov detectors.
  I discuss the robustness of the atmospheric neutrino oscillation
  hypothesis against the presence of Flavor Changing (FC) Non-Standard
  neutrino Interactions (NSI), generally expected in models of
  neutrino mass.  Atmospheric data strongly constrain FC-NSI in the
  \nm-\nt channel, while solar data can be explained by FC-NSI in the
  \ne-\nt channel, or, alternatively, by spin flavor precession.
  I illustrate how a neutrino factory offers a unique way to probe for
  FC-NSI and argue that a near-site detector is necessary in order to
  probe for leptonic mixing and CP violation.

\end{abstract}
\date{\today}

\section{ Solar Neutrinos \cite{sol,ssm,Kachelriess:2001sg,Gonzalez-Garcia:1999aj,Ahmad:2002jz,Ahmad:2002ka}}
\label{sec:solar-neutrinos}


For decades, the number of solar neutrinos detected in underground
experiments \cite{sol} has been less than expected from theories of
energy generation in the sun \cite{ssm}, suggesting that either the
understanding of the Sun was incomplete, or that the neutrinos were
changing from one type to another in transit from the core of the Sun
to the detector.
At the moment we have a conclusive proof for the latter possibility,
though we still can not pin down the exact neutrino conversion
mechanism implied.

Solar neutrinos have now been detected through the geochemical method
(the \ne + ${}^{37}$Cl $\to {}^{37}$Ar + $e^-$ reaction at the
Homestake experiment and the \ne + ${}^{71}$Ga $\to ^{71}$Ge + $e^-$
reaction at the Gallex, Sage and GNO experiments) and also directly
using Cherenkov techniques at Kamiokande and SuperKamiokande (SK) and
the Sudbury Neutrino Observatory (SNO).  At SK the reaction is $\nu_e
e$ scattering on water, while SNO detects neutrinos using 1000 tonnes
of heavy water (D2O) and is sensitive to the charged current (CC) and
neutral current (NC), as well as to the elastic scattering (ES)
reaction.

At present all experiments observe a deficit of 30 to 60 \% with an
energy dependence mainly due to the lower Chlorine rate.
The high statistics of SK after 1496 days of data--taking also
provides very useful information on the recoil electron energy
spectrum and this is presently well described by the flat hypothesis,
placing important restrictions on neutrino parameters.
Moreover, SK measures the zenith angle distribution (day/night effect)
which is sensitive to the effect of the Earth matter in the neutrino
propagation, observing no significant effect.
The absence of a clear hint of spectral distortion, day-night or
seasonal variation implies that, to a large extent the solar neutrino
problem rests heavily on the rate discrepancy and, from this point of
view, future experiments are most welcome.

\begin{figure}[htbp]
  \centering
\includegraphics[width=5cm,height=4cm]{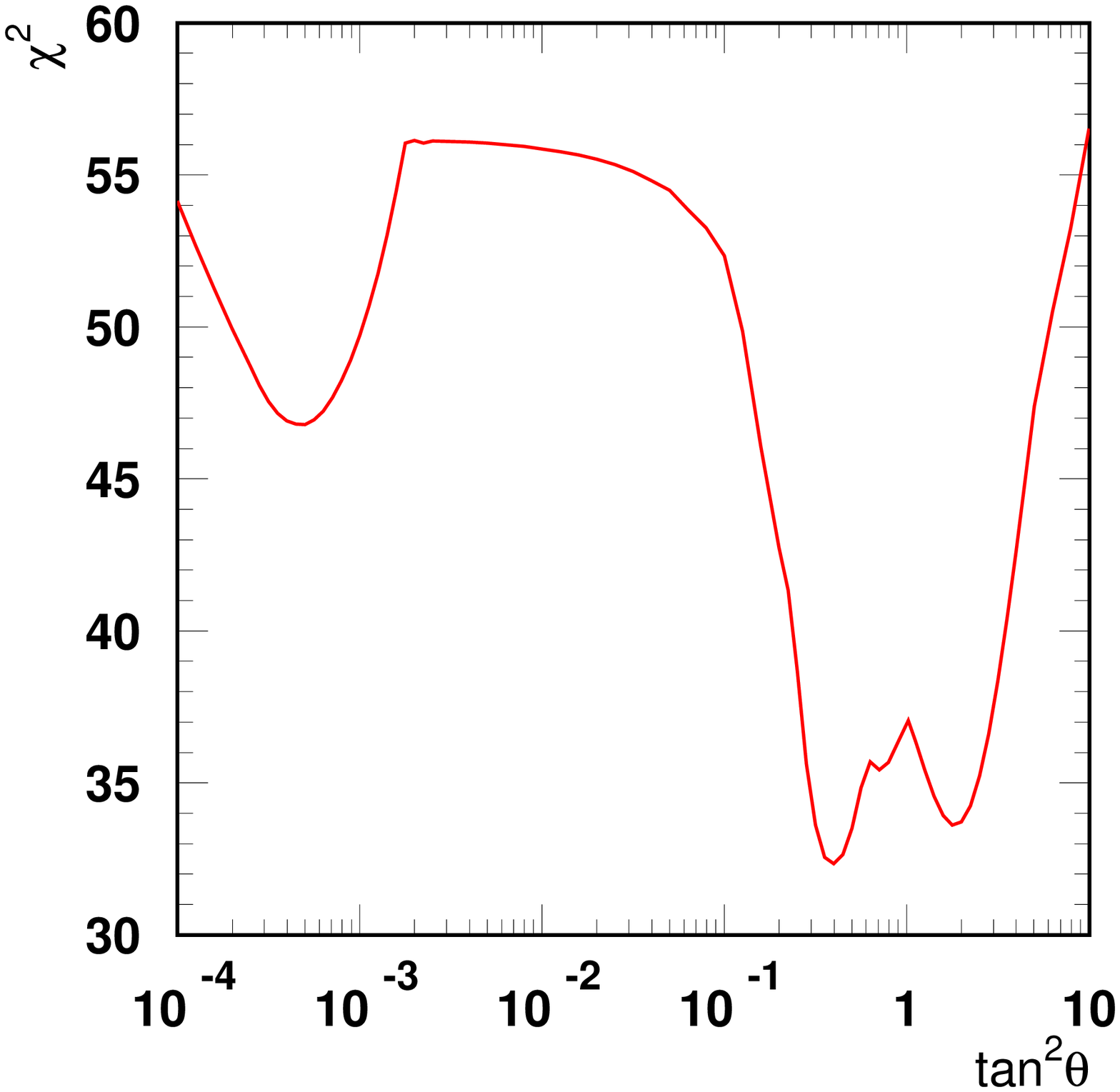}  
\includegraphics[width=5cm,height=4cm]{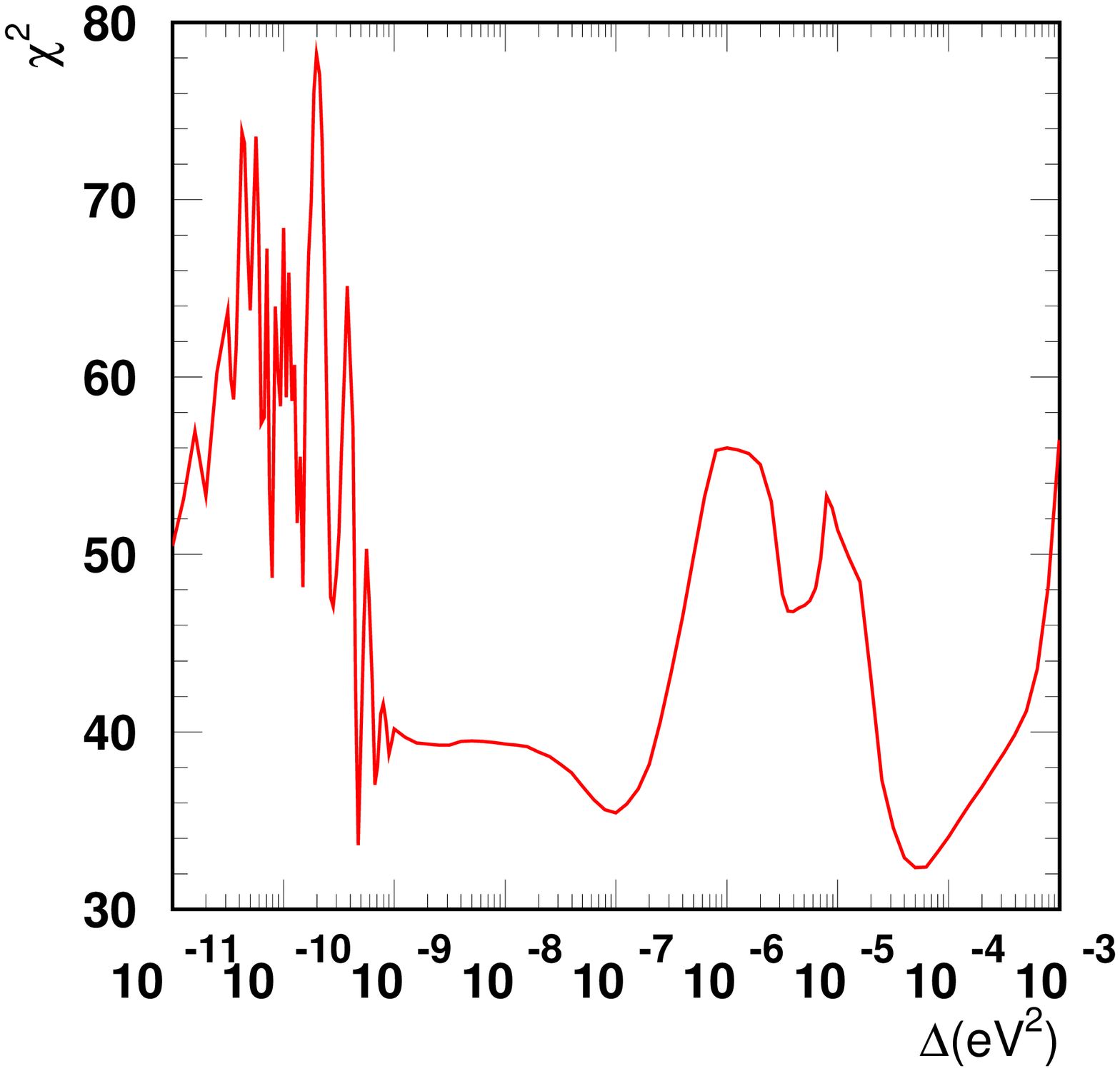}
  \caption{The solar $\chi^2$ profiles versus angle and $\Delta m^2$, 
    from \cite{Kachelriess:2001sg}, before
    \cite{Ahmad:2002jz,Ahmad:2002ka}.}
  \label{fig:solchi2}
\end{figure}

The global status of the interpretation of the solar neutrino data in
terms of active neutrino oscillations is illustrated in Fig.
\ref{fig:solchi2}.
This figure refers to the 1258 day SK data sample.  As can be seen the
best of the oscillation solutions is the Large Mixing Angle (LMA)
solution, confirming first pre-SNO hints obtained in ref.
\cite{Gonzalez-Garcia:1999aj}. This was derived solely on the basis of
the observed flatness of the recoil electron energy spectrum of SK,
highlighting its importance in shaping up what is now much more
strongly supported.

Solar neutrinos from the decay of $^8$B have been detected at SNO via
the CC reaction on deuterium and by the ES reaction.  Such CC reaction
is sensitive only to \ne's, while the ES reaction also has a small
sensitivity to \nm's and \nt's. SNO has recently observed also NC
neutrino interactions \cite{Ahmad:2002jz}.

\begin{figure}[htbp]
  \centering
\includegraphics[width=7cm,height=4.5cm]{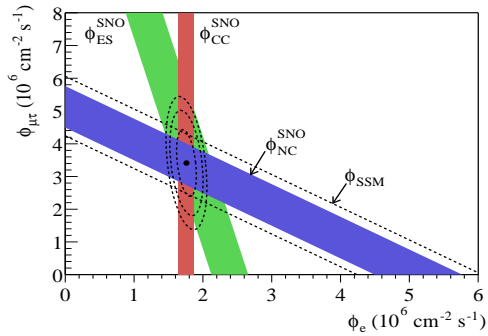}  
  \caption{SNO neutral current measurement \cite{Ahmad:2002jz}.}
  \label{fig:fig3sno}
\end{figure}

Using the standard ${}^{8}$B shape they measure the $\nu_{e}$
component of the ${}^{8}$B solar flux $\phi_{e}$ above 5 MeV as well
as the non-$\nu_{e}$ component $\phi_{\mu\tau}$. As shown in Fig.
\ref{fig:fig3sno} the latter is significantly above zero, providing
strong evidence for solar \ne conversion into another active neutrino
flavor.
This shows the flux of ${}^{8}$B solar neutrinos which are $\mu$ or
$\tau$ flavor versus the \ne flux derived from the three neutrino
reactions in SNO.  The diagonal bands show the total ${}^{8}$B flux as
predicted by the SSM~\cite{ssm} (dashed lines) and that measured with
the NC reaction in SNO (solid band).  The intercepts of these bands
with the axes represent the $\pm 1\sigma$ errors.  The bands intersect
at the fit values for $\phi_{e}$ and $\phi_{\mu\tau}$, indicating that
the combined flux results are consistent with neutrino flavor
conversion with no distortion in the ${}^{8}$B neutrino energy
spectrum.
They also find the total flux measured with the NC reaction to be
consistent with solar models.

SNO has also measured day and night solar neutrino energy spectra and
rates \cite{Ahmad:2002ka}.  They give an analysis in terms of
matter-enhanced oscillations of two active flavors showing how it
strongly favors the LMA solution. The update of global fits including
these latest data worsen the status of non-LMA and especially that of
non-active solutions, see ref.  \cite{ssm,solat02}. We eagerly await a
possible confirmation of the LMA hypothesis at KamLAND \cite{KamLAND}.
As we will see later there are also good active non-oscillation
neutrino conversion mechanisms based on flavor-changing neutrino NSI
or spin flavor precession that can accout well for the data.

\section{Atmospheric Neutrinos \cite{atm,atmfluxes,Gonzalez-Garcia:2000sq,Gonzalez-Garcia:2002mu,Fornengo:2000sr,Fornengo:2001pm}}
\label{sec:atmosph-neutr}

Atmospheric neutrinos are produced as the decay products in hadronic
showers from particles produced in collisions of cosmic rays with air
in the upper atmosphere. These have been observed in several
experiments \cite{atm}.
Although individual $\nu_\mu$ or $\nu_e$ fluxes are only known to
within $30\%$ accuracy, their ratio is predicted to within $5\%$ over
energies varying from 0.1~GeV to 100~GeV~\cite{atmfluxes}.
There has been a long-standing discrepancy between the predicted and
measured $\mu/e$ ratio of the muon ($\nu_\mu + \bar{\nu}_\mu$) over
the electron atmospheric neutrino flux ($\nu_e+\bar{\nu}_e$) both in
water Cherenkov experiments (Kamiokande, Super-Kamiokande and IMB) as
well as in the iron calorimeter Soudan2 experiment. Moreover a strong
\texttt{zenith-angle dependence} has been observed both in the sub-GeV
and multi-GeV energy range, but only for $\mu$--like events, the
zenith-angle distributions for the $e$--like being consistent with
expectation. Such clear deficit of neutrinos coming from below is very
suggestive of \nm oscillations.
Zenith-angle distributions have also been recorded for upward-going
muon events in SK and MACRO which are also consistent with the \nm
oscillation hypothesis.

\begin{figure}[htbp]
  \centering
\includegraphics[width=5.5cm,height=4cm]{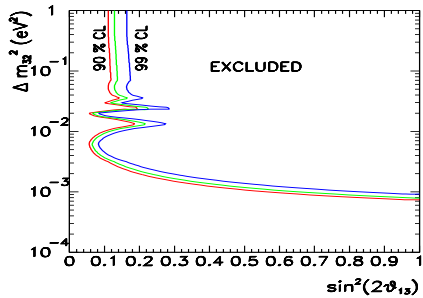}
  \caption{Region in  $\Delta{m}^2_{32}$ and $\sin^2(2\theta_{13})$  
  excluded by the Chooz reactor, from \cite{Gonzalez-Garcia:2000sq}.}
  \label{fig:chooz}
\end{figure}

\begin{figure}[htbp]
  \centering
\includegraphics[width=5.5cm,height=4cm]{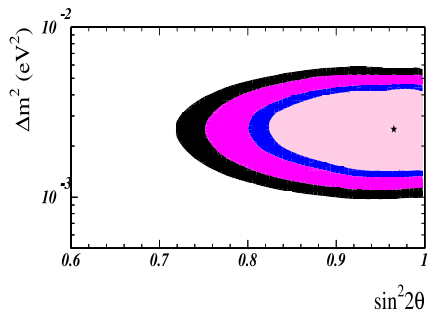}  
  \caption{Atmospheric oscillation parameters from global analysis in 
    \cite{Fornengo:2001pm}.}
    \label{fig:atm}
\end{figure}

Of all laboratory searches for neutrino oscillation, reactors provide
the highest sensitivity in $\Delta{m}^2_{32}$ and thus are most
relevant to confront with the data from underground experiments.  The
restrictions on $\Delta{m}^2_{32}$ and $\sin^2(2\theta_{13})$ that
follow from the non observation of oscillations at the Chooz reactor
experiment are shown in Fig.~\ref{fig:chooz}.
The curves represent the 90, 95 and 99\% CL excluded region defined
with 2 d.~o.~f. for comparison with the Chooz published results.
For large $\Delta{m}^2_{32}$ this gives a stringent limit on
$\sin^2(2\theta_{13})$, but not for low $\Delta{m}^2_{32}$ values.
After combining with the full body of atmospheric neutrino data, one
finds a stringent bound $\sin^2(\theta_{13}) < 0.045$ at $99\%$ CL
\cite{Gonzalez-Garcia:2002mu}.  Neglecting $\theta_{13}$ we display in
Fig.~\ref{fig:atm} the allowed regions of atmospheric oscillation
parameters $\Delta{m}^2_{32}$--$\sin^2(2\theta_{23})$ obtained in a
global analysis.  For previous analyzes, see \cite{Fornengo:2000sr}
and references therein.

\section{Supernova Neutrinos 
  \cite{Kachelriess:2000fe,Minakata:2001cd}.}
\label{sec:supernova-neutrinos}

The large-angle neutrino oscillation between the electron neutrino \ne
favored by current solar neutrino data must be properly taken into
account in the interpretation of the SN neutrino data.
Indeed, it will induce a significant deformation of the energy
spectrum of neutrinos coming from a supernova (SN) explosion and on
this basis it has been argued \cite{snperm} to be at odds with the
historical observation of neutrinos from SN1987 \cite{Hirata:1987hu}.

The impact of $\bar\nu_e \leftrightarrow \bar\nu_{\mu,\tau}$ neutrino
oscillations on the observed $\bar\nu_e$ signal of supernova SN 1987A
can be studied by performing a maximum-likelihood analysis. The fit
parameters are the released binding energy $\Eb$ and the average
neutrino energy $\Ee$.  It was found that $\bar\nu_e \leftrightarrow
\bar\nu_{\mu,\tau}$ oscillations with large mixing angles have lower
best-fit values for $\Ee$ than small-mixing angle (SMA) oscillations.
Moreover, the inferred value of $\Ee$ is already lower than found in
simulations in the SMA or no-oscillation case. This apparent conflict
has been interpreted as evidence against the large mixing oscillation
solutions to the solar neutrino problem.
In order to quantify the degree to which the SN data favor
no-oscillations over the large mixing solutions their likelihood
ratios were used as well as a Kolmogorov-Smirnov test. The result was
that within the range of SN parameters predicted by simulations there
are regions where the LMA solution is either only marginally
disfavored or even favored over the SMA case
\cite{Kachelriess:2000fe}.
This implies that the LMA solution is not in conflict with the current
understanding of SN physics. In contrast, vacuum oscillation and LOW
solutions may be excluded at the $4\sigma$ level for most of the SN
parameter ranges found in simulations. The reason for the difference
amongst the large mixing solutions has to do with the role of \ne
regeneration in the Earth matter.

One can go a step further, namely to perform a combined likelihood
analysis of the observed SN1987A neutrino signal and of the solar
neutrino data \cite{Kachelriess:2001sg}.
In the calculation of neutrino survival and conversion probabilities a
power-law $1/r^3$ profile typical of supernova envelopes was used and
the level-crossing probability was determined analytically in a form
valid for all mixing angles \cite{Kachelriess:2001bs}.
It has been found that, although the spectrum swap argument is
sufficient to rule out vacuum-type solutions for most reasonable
choices of astrophysics parameters, the LOW solution may still be
acceptable.
On the other hand it was found that the LMA solution can easily
survive as the best overall solution, although its size is typically
reduced when compared to solar-only fits \cite{Kachelriess:2001sg}.
How about a future supernova?

\begin{figure}[htbp]
  \centering
\includegraphics[width=5cm,height=5cm]{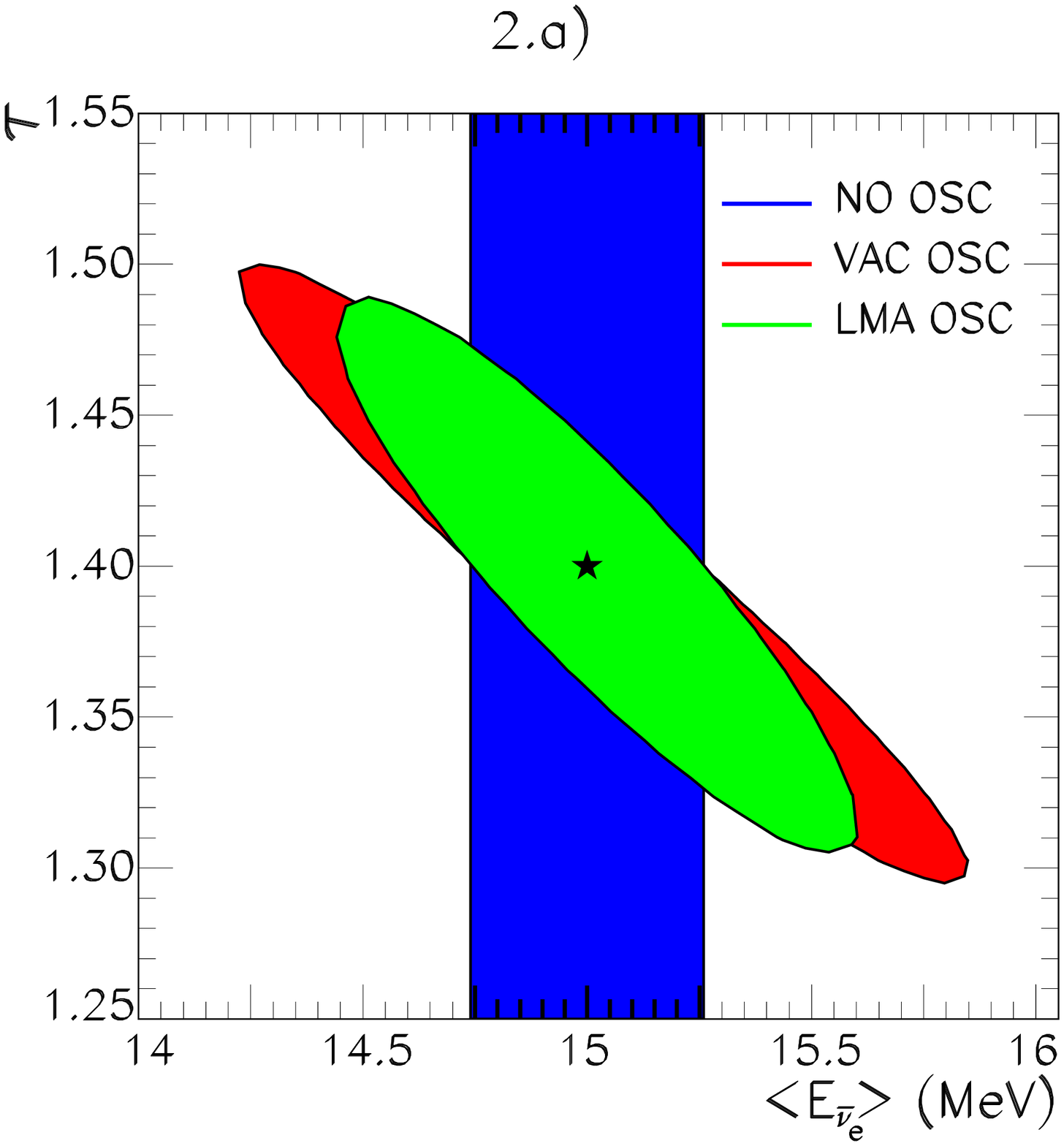}
\includegraphics[width=5cm,height=5cm]{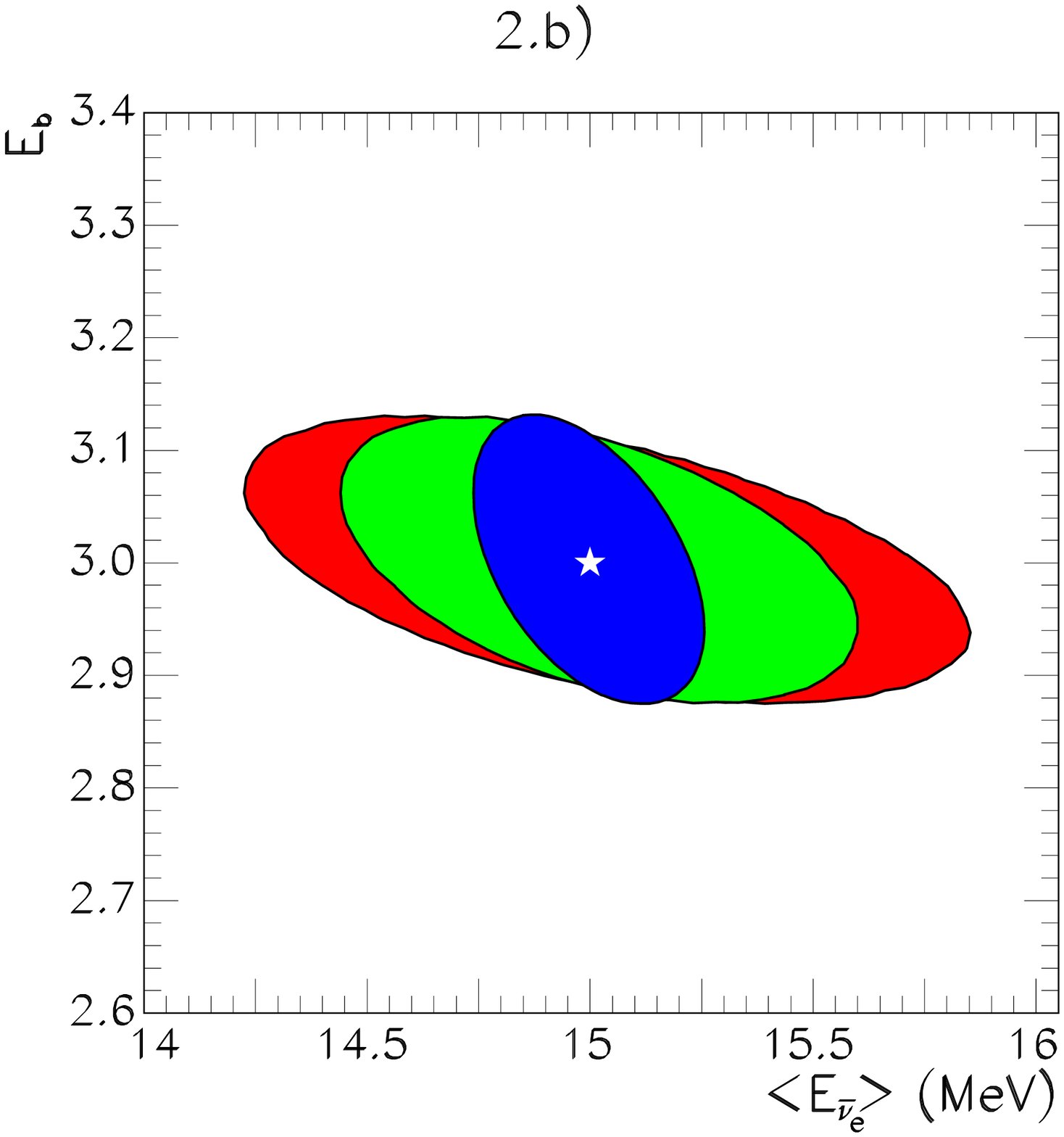}  
  \caption{LMA \ne oscillations allow a determination of 
SN parameters \cite{Minakata:2001cd}.}
  \label{fig:minak}
\end{figure}

The significant distortion of the energy spectrum of neutrinos coming
from a supernova (SN) explosion expected by active LMA \ne
oscillations can be used to extract some of the properties of SN
driven by gravitational collapse.
In ref. \cite{Minakata:2001cd} a simple but powerful method was
proposed to extract the original temperatures of both anti-electron
and muon/tau neutrinos at the neutrinosphere from the observation of
anti-electron neutrinos at a large water Cherenkov detector such as
Super-Kamiokande. Fig. \ref{fig:minak} illustrates the potential of
the method in determining intrinsic SN parameters.
As can be seen one obtains a rather good determination of the
important astrophysical parameter $\tau \equiv {\langle
  E_{\bar{\nu}_\mu}\rangle}/{\langle E_{\bar{\nu}_e}\rangle }$
characterizing the non-electron flavor SN neutrino spectra. The result
is rather robust in the sense that it is not too sensitive to
astrophysical assumptions such as energy equipartition.
Prospects are substantially better for the projected megaton-class SK
successor, HyperKAMIOKANDE.

\section{Basic Neutrino Theory \cite{Weinberg:uk,seesaw,Mohapatra:1979ia,Schechter:1980gr,Schechter:1980gk,Chikashige:1980ui,Schechter:1981cv,Schechter:1981bd,Pal:1981rm,Schechter:1981hw}}
\label{sec:basic-neutr-theory}

The basic gauge theoretical aspects of neutrinos have been developed
in the early eighties
\cite{Weinberg:uk,seesaw,Mohapatra:1979ia,Schechter:1980gr,Schechter:1980gk,Chikashige:1980ui,Schechter:1981cv,Schechter:1981bd,Pal:1981rm,Schechter:1981hw}.
The other basic step was the formulation of the MSW effect itself
\cite{Wolfenstein:1977ue}. With both ingredients one can describe all
the exciting experimental developments of the past few years
\cite{Bilenky:1996rw}.
In order to make sense of present neutrino data through neutrino
oscillations, neutrinos should have mass. In analogy with the
Kobayashi-Maskawa (KM) quark mixing matrix \cite{Kobayashi:fv}, let
$K$ denote the lepton mixing matrix.  The simplest form for $K$ in a
gauge theory of the weak interaction is characterized by
\begin{itemize}
\item 
the atmospheric angle $\theta_{23} \equiv \theta_A$
\item 
 the solar
  angle $\theta_{12} \equiv \theta_S$ 
\item 
the reactor angle
  $\theta_{13} \equiv \theta_R$
\item
1 KM-like CP phase $\phi_{13}$
\item
 2 extra (Majorana) phases $\phi_1, \phi_2$ \cite{Schechter:1980gr}
\end{itemize}
The ``Majorana'' phases are physical \cite{Schechter:1980gk}. Although
they do not show up in ordinary ($\Delta L=0$) oscillations, they
affect $\Delta L=2$ processes. Unfortunately, however, the effect is
tiny due to helicity suppression, reflecting the V-A nature of the
weak interaction \cite{Schechter:1980gk}.
Also the effect of the ``Dirac'' CP violation due to $\phi_{13}$ is
small since it disappears as $\theta_{13} \to 0$
\cite{Schechter:1979bn}.
A new generation of long-baseline neutrino oscillation experiments
using a neutrino beam from the decay of muons in a storage ring is
being considered~\cite{nuFacPapers}.  Studying $\theta_{13}$ and the
resulting leptonic CP violating effects provides a challenge for such
so-called \texttt{neutrino factories}.

The structure of charged and neutral current weak interactions is more
complex in models of neutrino mass containing \21 singlet
leptons~\cite{Schechter:1980gr}: the CC is characterized by a
rectangular (effectively non-unitary) matrix, while the NC is
described by a projective matrix with non-diagonal entries coupling
different mass eigenstate neutrinos with each other. Similarly, if
there are light sterile neutrinos, the nontrivial structure of $K$
must be taken into account.

\begin{figure}[htbp]
  \centering
\includegraphics[width=4cm,height=2.6cm]{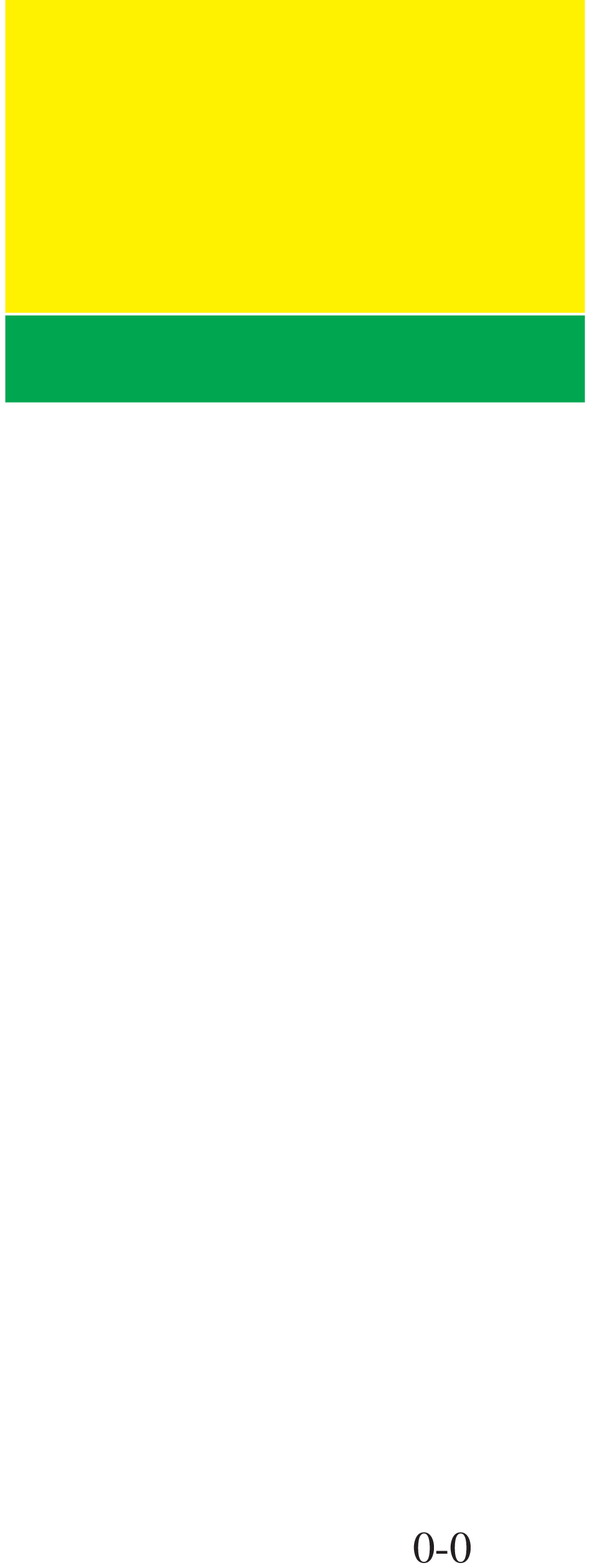}
\includegraphics[width=4cm,height=2.6cm]{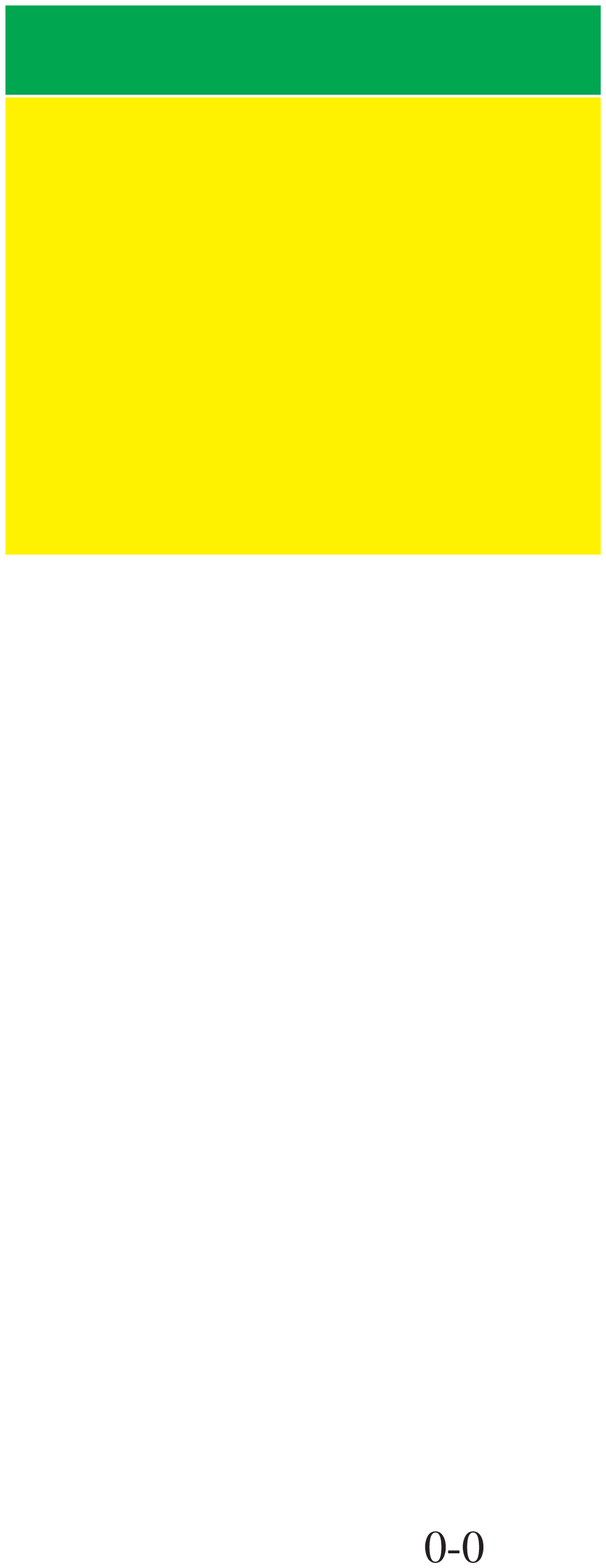}   
  \caption{Normal and inverse-hierarchical neutrino masses. The 
wide and narrow  bands represent the atmospheric  and solar 
mass splittings, respectively.}
  \label{fig:3nu-patterns}
\end{figure}

\begin{figure}[htbp]
  \centering
\includegraphics[width=5cm,height=4.cm]{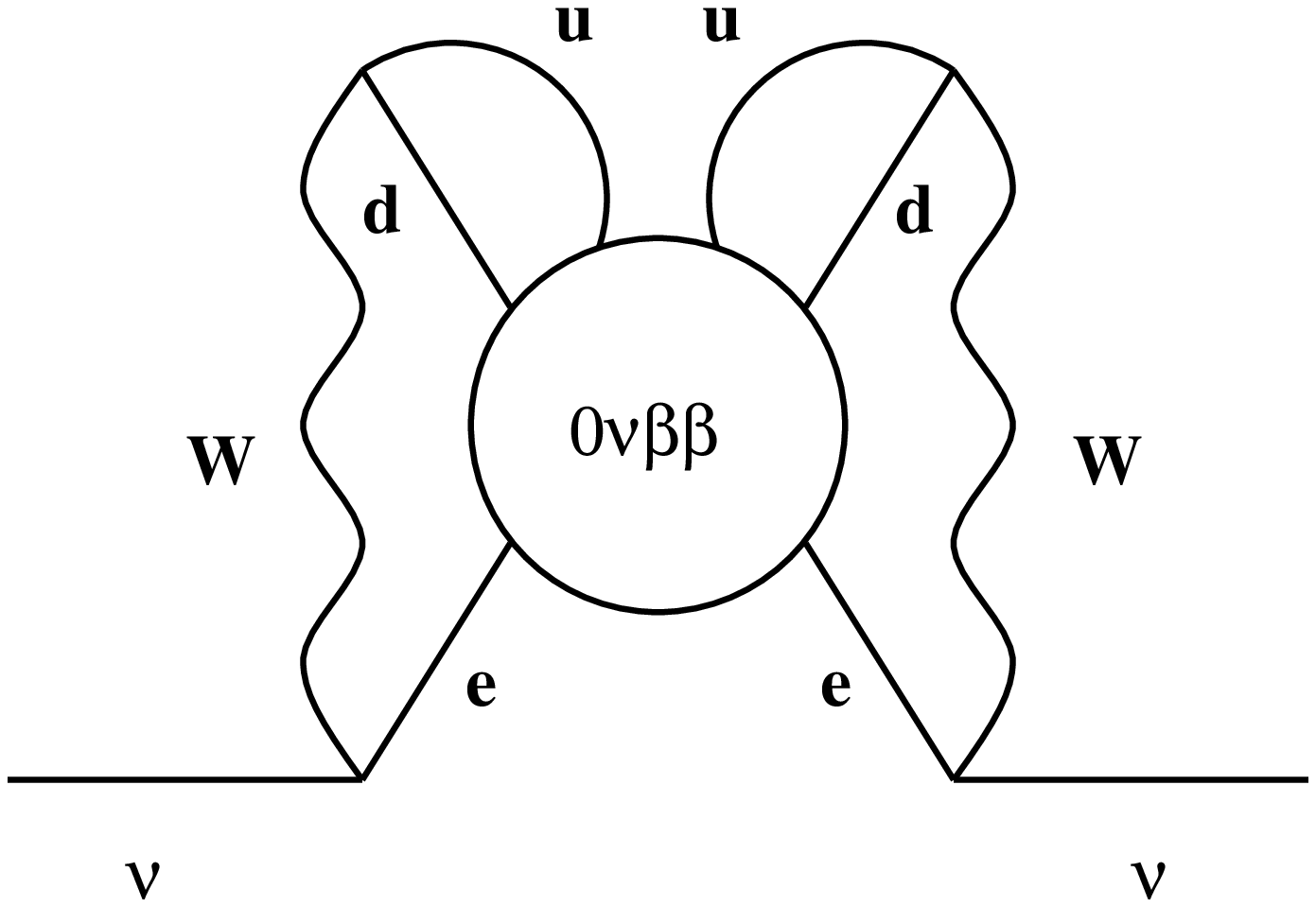}  
  \caption{The black-box \nbb argument, from \cite{Schechter:1981bd}.}
 \label{fig:bbox} 
\end{figure}

Current solar and atmospheric data can be understood well with the
usual three massive neutrino flavors. The information on the neutrino
oscillation parameters may be summarized by saying that $\theta_{23}$
is nearly maximal, $\theta_{12}$ is large, but non-maximal, and the
reactor angle $\theta_{13}$ is rather small.  In contrast there is at
the moment no information on the CP phases.
It is clear from the above that the atmospheric mass splitting is much
larger than the solar one, $\Delta {m_{atm}^2} \gg \Delta
{m_\odot^2}$. Such hierarchical pattern can be realized in two ways,
illustrated in Fig.~\ref{fig:3nu-patterns}.
Last, but not least, the oscillation data are sensitive only to
neutrino mass splittings, not to the absolute scale of neutrino mass.
Nor do they have any bearing on the fundamental issue of whether
neutrinos are Dirac or Majorana particles. This brings us to the
significance of the \nbb decay \cite{Morales:1998hu} in deciding the
nature of neutrinos.  The connection between the two is given by the
\texttt{black-box theorem} which states that, in a ``natural'' gauge
theory, whatever the mechanism for inducing \nbb is, it is bound to
also yield a Majorana neutrino mass at some level, and vice-versa, as
illustrated by Fig.  \ref{fig:bbox}.
Quantifying the implications of the black-box argument is a
model-dependent enterprise, but can be done once a particular model
is specified. It serves at least to highlight the importance of
confirming or refuting the present hint
\cite{Klapdor-Kleingrothaus:2001ke,Aalseth:2002dt} in a more sensitive
experiment such as GENIUS \cite{Klapdor-Kleingrothaus:1999hk}.

Can one predict neutrino properties from first principles?  Such
important enterprise is unfortunately difficult, for at least three
basic reasons.
First the underlying \texttt{scale} is unknown, models using anything
between the Planck or String scale, through the \e6 or \10 GUT scale,
or an intermediate scale associated with Peccei-Quinn or Left-Right
symmetry, all the way down to the weak \321 scale itself.
Second, the \texttt{mechanism} is unknown, and there are both tree
level, like the so-called seesaw schemes
\cite{seesaw,Schechter:1980gr,Chikashige:1980ui,Schechter:1981cv} as
well as radiative \cite{Zee:1980ai} or hybrid alternatives.
Similarly the status of B-L (baryon minus lepton number) symmetry is
unknown, it may either be a broken gauge symmetry, such as in the
left-right or \10 schemes, or simply a global ungauged symmetry, in
which case the corresponding Goldstone boson -- called majoron -- is
physical \cite{Chikashige:1980ui,Schechter:1981cv}.
Last, but not least, there is \texttt{no theory of flavour}, although
many Yukawa textures have been suggested which would arise from
suitable family symmetries.

Renormalizable neutrino masses are forbidden by the Standard Model
(SM), given its restricted field content and symmetries.  The simplest
and most basic way to add neutrino masses is to regard the SM only as
the effective low-energy limit of some, unknown, higher energy theory,
which may well contain an L-violating dimension-five term
\cite{Weinberg:uk}
\begin{equation}
{\cal L}_5= \frac{\lambda_{ij}}{M_5}(\ell_i \phi)(\ell_j \phi),
\label{5-dim}
\end{equation} 
involving the lepton doublet fields $\ell_i$ ($i,j=e,\mu,\tau$), and
the Higgs doublet field $\phi$. 
\begin{figure}[htbp]
  \centering
\includegraphics[width=4.5cm,height=3cm]{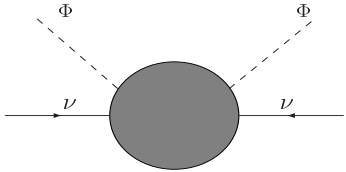}  
  \caption{Neutrino masses from dimension-five operator \cite{Weinberg:uk}.}
  \label{fig:d-5}
\end{figure}
Here $M_5$ is an arbitrary mass scale, related to the unknown high
energy theory of which the SM is an effective theory and
$\lambda_{ij}$ are dimensionless couplings.
After electroweak symmetry breaking, eq. (\ref{5-dim}) leads to
neutrino Majorana masses $m_{\nu}^{ij}$ given by
\begin{equation}
m^{\nu}_{ij}=\frac{\lambda_{ij}v^2}{M_5}.
\label{mass}
\end{equation}
where $v=174$~GeV is the Higgs vacuum expectation value (vev). Note
that eq.~(\ref{mass}) explicitly violates lepton number symmetry.  

One way to realize this picture is in terms of the so-called seesaw
scenario for neutrino
mass~\cite{seesaw,Mohapatra:1979ia,Schechter:1980gr,Chikashige:1980ui,Schechter:1981cv},
where this operator is induced either through the exchange of heavy
\321 singlet neutrinos or extra scalar bosons related, say, with the
breaking of extended left-right-type gauge symmetries.  The first
challenge for unification seesaw schemes is that the smallness of the
quark KM mixing angles typically suggests that leptonic mixing angles
should also be small.  However this is certainly possible to avoid
owing to the lack of predictivity. Indeed, despite its attractiveness
and the existence of many papers claiming to have ``predictive''
seesaw models of neutrino mass and mixing \cite{seesawnupapers}, such
models at most \texttt{post-dict} the observed angles and splittings
and it is hardly possible to obtain ``smoking-gun'' tests of their
validity.
  
\section{Neutrino Models}
  
Notwithstanding the limitations of theory in predicting neutrino
properties from first principles, I will mention some recent examples
of predictive theoretical schemes for accounting for the neutrino
anomalies.  Given the vast literature \cite{seesawnupapers,giunti}, I
must confine here to a personal selection of recent neutrino models.

\subsection{Neutrino Mass Unification~\cite{Chankowski:2000fp}}
\label{sec:neutr-mass-unif}

\begin{figure}[htbp]
  \centering
\includegraphics[width=6cm,height=3cm]{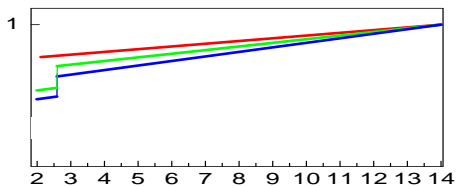}  
  \caption{Artists' view of  neutrino masses arising from a 
    common seed (1 eV) at a large neutrino unification scale
    $M_X \sim 10^{14}$ GeV~\cite{Chankowski:2000fp}.}
  \label{fig:pok}
\end{figure}

The idea of neutrino mass unification is inspired by the paradigm of
unification of fundamental interactions. This ``minimalistic'' ansatz
postulates that the neutrino mass and mixings observed at low energies
take a very simple form at some high energy \texttt{neutrino
  unification} scale $M_X$.  The model adds to the basic Lagrangian
the dimension--five operator in eq.~(\ref{5-dim}) assumed to be
characterized at the scale $M_5 = M_X$ by a single real
dimension--less parameter $\lambda_0$.
In this picture neutrino masses arise from a common seed at $M_X$,
while their splittings are induced by renormalization effects, as
illustrated in Fig. \ref{fig:pok}.
This model is inconsistent with small solar mixing, thus it
``predicts'' a large, though non-maximal solar angle, according to the
formula $$\Delta m^2_{S}~\approx~2 \Delta m^2_{A}
\left[\cos~2~\theta_{S} + 2~\sin^2 \theta_{R} \right],$$
where $\Delta m^2_{S}$ and $\Delta m^2_{A}$ are the solar and
atmospheric splittings while $\theta_{S}$ and $\theta_{R}$ are the
solar and reactor angles, respectively.
Neutrinos can lie in the eV range and thus be seen in tritium beta
decays \cite{Osipowicz:2001sq} as well as contribute to the hot dark
matter \cite{Elgaroy:2002bi}.  In contrast to other mechanisms leading
to quasi-degenerate neutrinos \cite{Caldwell:kn,DEG} which may be
unnatural \cite{DEGNAT}, here this is not the case because opposite
neutrino CP signs
\cite{Schechter:1981hw,Wolfenstein:1981rk,Valle:1982yw} suppress the
rate for neutrinoless double beta decay.
Taken at face value, an analysis including the most recent NC SNO data
\cite{ssm} suggests that the preferred solution in this case will be
in terms of vacuum oscillations, instead of LMA, which as we saw is
disfavored by SN1987A.  In order to evaluate more carefully our
prediction a three-neutrino global fit of all neutrino data, similar
to that in \cite{Gonzalez-Garcia:2000sq}, would be required.

\subsection{Minimalistic Seesaw Plus Gravity~\cite{deGouvea:2001jp}}
\label{sec:minimalism}

\begin{figure}[htbp]
  \centering
\includegraphics[width=4cm,height=3cm]{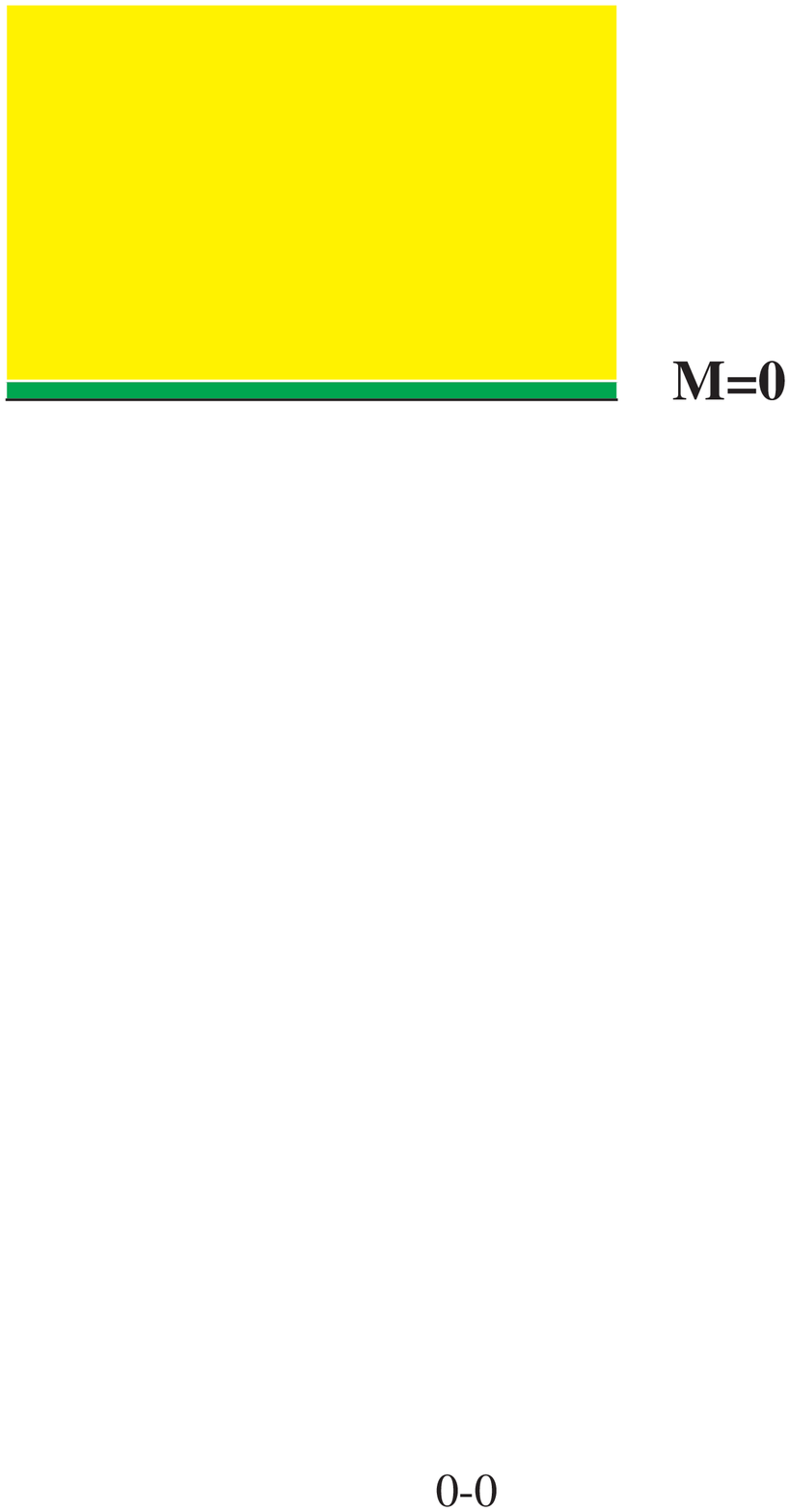}  
\includegraphics[width=4cm,height=3cm]{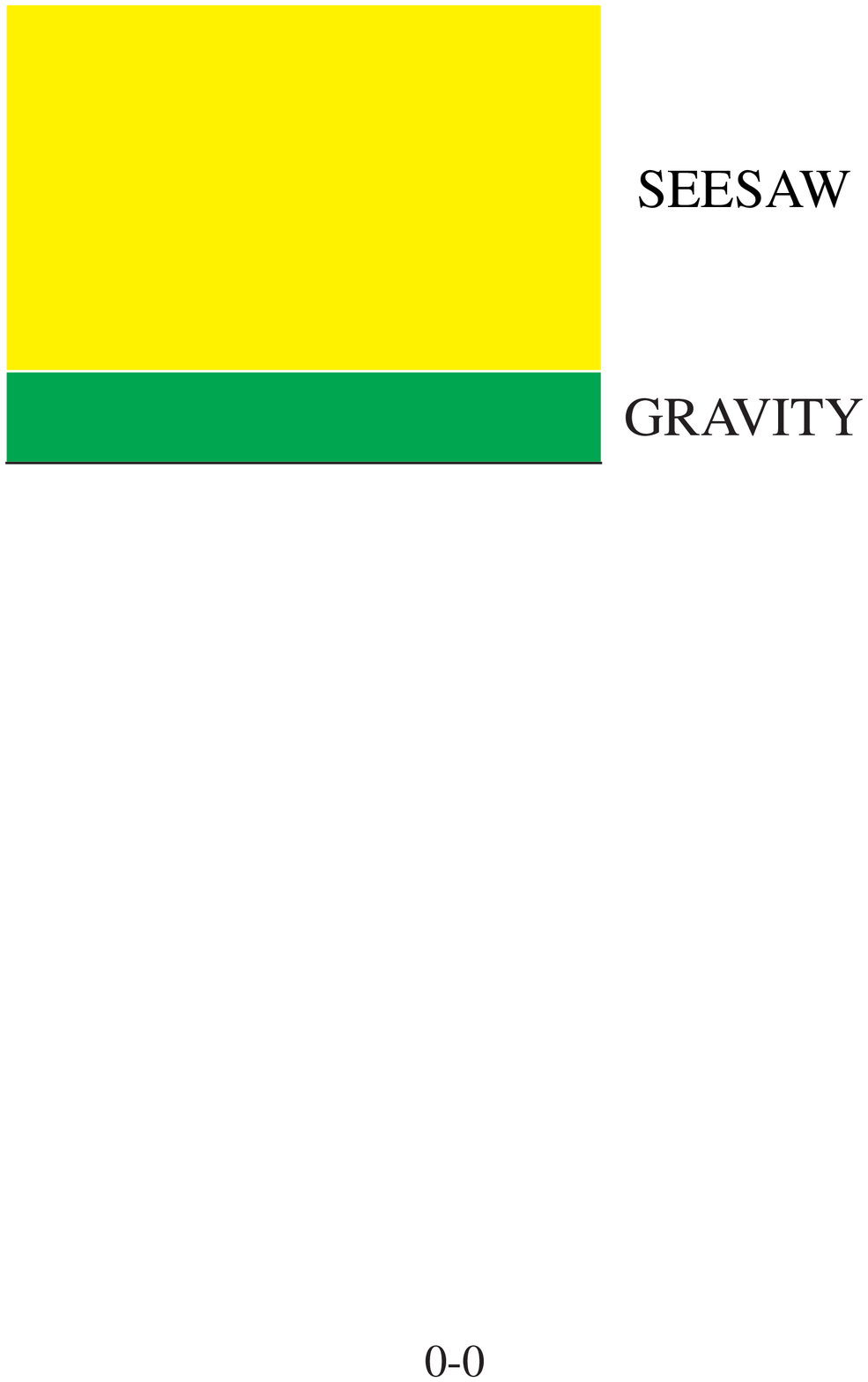}
  \caption{Minimalistic neutrino masses. The degeneracy of the 
    massless states on the left is lifted by gravitational effects,
    illustrated in the right panel.}
  \label{fig:gouv}
\end{figure}

The idea here rests on the observation that gravity alone can account
for the solar neutrino data by Planck-mass L-violating effects, while
the atmospheric neutrino anomaly can be due to the existence of a
single right-handed neutrino, instead of
three~\cite{Schechter:1979bn}. One assumes it to be at an intermediate
mass scale between $10^9$ GeV and $10^{14}$ GeV. To first
approximation only one of the three flavor neutrinos acquires mass
\textit{a la seesaw}, identified as the atmospheric scale. The other
two neutrinos remain massless, as illustrated in Fig. \ref{fig:gouv}
(left panel). Their degeneracy is lifted only by tiny gravitational
effects (right panel) due to the dimension--five operator.
Such simple scheme solves the current solar and atmospheric neutrino
puzzles with probably the smallest amount of beyond-the-Standard-Model
ingredients. Note that although the neutrino mixing angles are not
exactly predicted, they can be large, which agrees well with the
current experimental situation.  However it requires the vacuum or
just-so solution to the solar neutrino problem, not currently favored.
One test for the model is the search for anomalous seasonal effects at
Borexino \cite{bxino}.

\subsection{Supersymmetry As the Origin of Neutrino Mass 
  \cite{Hirsch:2000ef,Mira:2000gg}}
\label{sec:supersymm-as-orig}

The suggestion that neutrino masses may be genuinely supersymmetric is
rather old \cite{rpnu-spo1,rpnu-exp}. One simply postulates that they
arise as a result of the violation of R parity (RPV),
$$R_{p}=(-1)^{3B+L+2S}$$
where $S, B, L$ denote spin, baryon and total
lepton number, respectively. Standard model particles (including Higgs
scalars) are RP-even, while their superpartners are odd.  R parity
acts as a selection rule, according to which supersymmetric particles
can only be pair-produced, the lightest \texttt{sparticle} being
stable.  These properties have been taken to be the basis of most
searches of supersymmetric particles in the laboratory. However, R
parity is by no means a sacred symmetry and its violation could
substantially affect the expected SUSY signatures while endowing
neutrinos with mass.

\begin{figure}[htbp]
  \centering
\includegraphics[width=8cm,height=5cm]{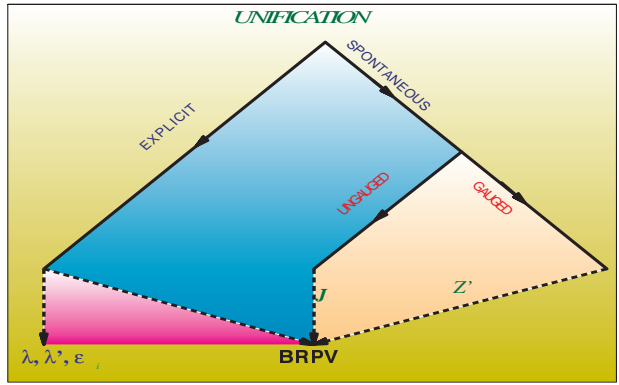}  
  \caption{Ways to break R parity.}
  \label{fig:rpcase1}
\end{figure}

There are several ways to realize this as illustrated in Fig.
\ref{fig:rpcase1}.
The violation of R--parity can arise explicitly \cite{rpnu-exp} as a
as a residual effect from unification that filters down to low
energies, or simply as an electroweak phenomenon generated due to
non-zero \21 singlet sneutrino vevs~\cite{rpnu-spo2}. The latter may
be realized either with gauged or ungauged lepton number, in the
latter case there is a majoron.

Barring the existence of special symmetries, when R--parity breaks
explicitly there can be both bi-linear and tri-linear terms in the
low-energy superpotential
$$W= W_{MSSM} + \mu_\alpha \ell_\alpha H_u + {\rm trilinears}.$$
However there may exist symmetries which simplify the otherwise very
complex structure of explicit RPV terms at low-energy, characterized
by too many new independent coupling constants.
An example of this possibility arises within the context of theories
with \texttt{anomalous $U(1)_H$ horizontal symmetries}. It has been
shown~\cite{Mira:2000gg} how one can choose the symmetry in such a way
that all 45 trilinear R-parity violating couplings are forbidden by
holomorphy. The symmetry also provides a common origin for the
supersymmetric $\mu$ term and the solution of the solar and
atmospheric neutrino anomalies, potentially explaining also the
required hierarchy between these.

Alternatively, if R parity is violated spontaneously \cite{rpnu-spo2}
necessarily one obtains the bi-linear RPV model (BRPV) as the
effective low-energy theory. Thus the BRPV model is the ``common
denominator'' of the three most attractive routes in Fig.
\ref{fig:rpcase1}, ``closed'' under the renormalization group. It also
constitutes the simplest and most predictive extension of the MSSM
which renders a very systematic investigation of R parity violation
effects not only in neutrino physics but also for supersymmetry
phenomenology at collider experiments.

\begin{figure}[htbp]
  \centering
\includegraphics[width=4cm,height=3cm]{solatm0.ps}
\includegraphics[width=4cm,height=3cm]{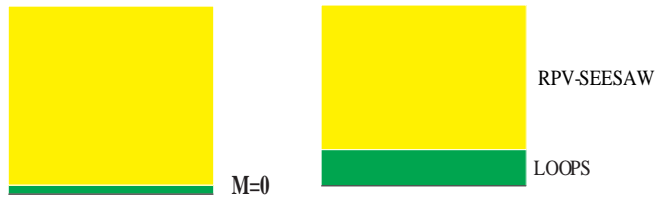}  
  \caption{Neutrino masses from RPV supersymmetry. 
    The degeneracy of the massless states on the left is lifted by
    supersymmetric loop effects, illustrated in the right panel.}
  \label{fig:solatrpsusy}
\end{figure}

On general grounds the spectrum of the BRPV model is expected to be
hierarchical, since only one of the three neutrinos picks up mass at
the tree level, by mixing with neutralinos, as indicated in Fig.
\ref{fig:solatrpsusy} (left panel).
Their degeneracy is lifted only by calculable loop effects (right
panel) \cite{Hirsch:2000ef} due to supersymmetric particle
exchanges~\cite{Diaz:1998xc}.
Both the tree level atmospheric scale as well as the more complex
loop-induced solar mass scale depend quadratically on RPV parameters.
On the other hand neutrino mixing angles are given as RPV parameter
ratios.  For example the left panel in Fig. \ref{fig:atmtest} shows
the correlation between $\tan^2_{\theta_A}$ with the
$\Lambda_2/\Lambda_3$ ratio. Similar correlations exist expressing the
solar and reactor angles as suitable RPV ratios \cite{Hirsch:2000ef}.

In summary, the BRPV model provides a powerful dynamical scheme for
neutrino masses which can account for the observed atmospheric and
solar neutrino anomalies in terms large angle neutrino oscillations,
as indicated by current data.

\section{Neutrino Mixing Tests at Colliders 
\cite{Porod:2000hv,Restrepo:2001me}}
\label{sec:neutr-mixing-tests}

Despite the smallness of neutrino masses indicated by the oscillation
interpretation of current neutrino data, R-parity violation can be
observable at high--energy collider experiments such as the
LHC~\cite{Bartl:2001yh}.
The first and most direct manifestation of RPV is the decay of the
lightest supersymmetric particle (LSP).  In many supergravity
scenaria, such as the BRPV model, the LSP is typically the lightest
neutralino and its mass could potentially be explored at LHC and other
planned colliders.
Since R parity is violated the LSP decays as
$$\chi^0 \to f_1 + \bar{f_2} + \ell$$
where $f_i$ denotes a quark,
lepton or neutrino and $\ell$ denotes either a charged lepton or a
neutrino.
\begin{figure}[htbp]
  \centering
\includegraphics[width=6cm,height=4.5cm]{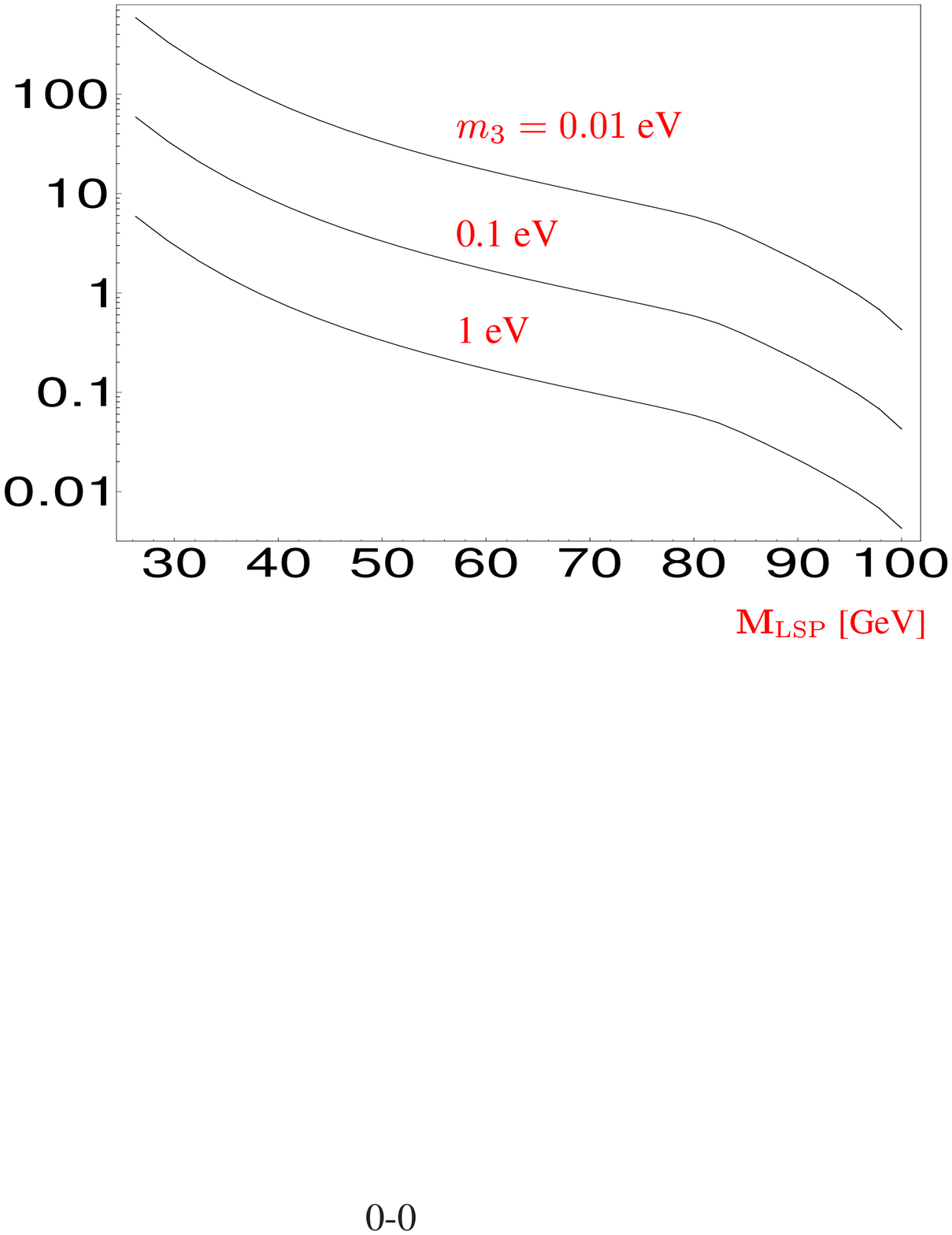}  
  \caption{LSP decay length (in cm) vs its mass (in GeV) in the BRPV model,
    from \cite{Bartl:2001yh}.}
  \label{fig:lspdecaylength}
\end{figure}
In Fig. \ref{fig:lspdecaylength} we show the typical LSP decay length
expected in the BRPV model. Clearly the decay length is short enough
to be studied at the LHC. Moreover the decay products are mostly
visible, with the main decay channel being semi-leptonic, into two
b-jets plus missing transverse momentum.

\begin{figure}[htbp]
  \centering
\includegraphics[width=4cm,height=3cm]{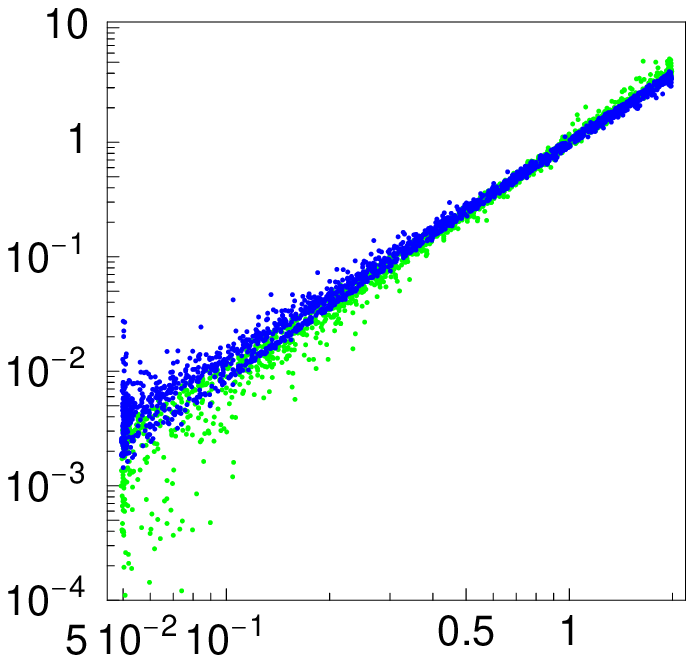}
\includegraphics[width=4cm,height=3cm]{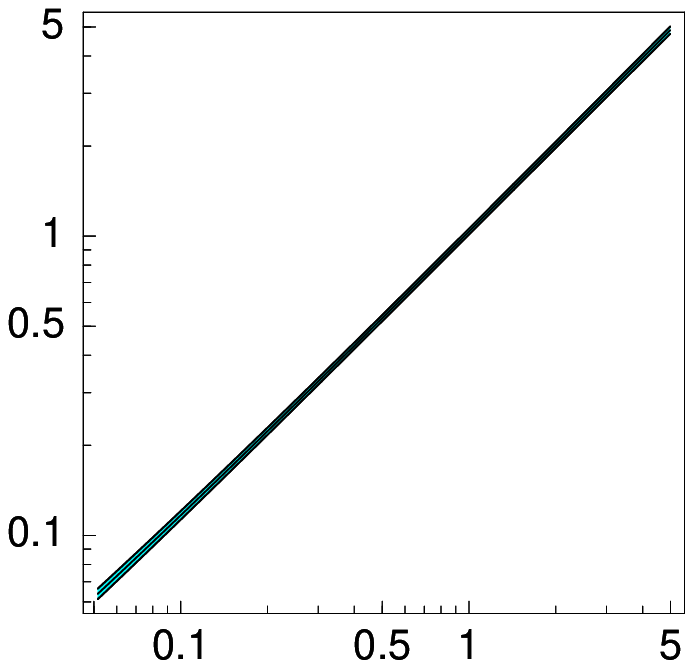}  
  \caption{Testing the atmospheric angle in LSP decays, 
from \cite{Porod:2000hv}.}
  \label{fig:atmtest}
\end{figure}

Most importantly, the BRPV model provides an unambiguous test of the
mixing angles involved in the neutrino anomalies at accelerators, by
measuring the decay patterns of the lightest neutralino
\cite{Porod:2000hv}. For example, in the right panel of Fig.
\ref{fig:atmtest} I illustrate how the ratio of semileptonic
neutralino decay branching ratios to muons and tau leptons is
\texttt{predicted} by the measured value of the atmospheric mixing
angle indicated by atmospheric data.
Similar considerations hold if the LSP in not the lightest neutralino,
but some other supersymmetric particles. For example when the stop
quark is the LSP, similar correlations exist between the stop decay
branching ratios and the solar angle \cite{Restrepo:2001me}.

\bigskip

\section{Life Beyond Oscillations? }
\label{sec:life-beyond-oscill}

Massive neutrinos typically have Non-Standard Interactions (NSI),
which can be both flavour-changing (FC) and non-universal (NU)
\cite{Valle:1990pk}.

\texttt{Gauge-type NSI} are defined as coming from a nontrivial
structure of CC and NC weak interactions. They arise in seesaw-type
schemes of neutrino mass due to the rectangular (effectively
non-unitary) nature of the charged current lepton mixing matrix and
correspondingly non-trivial neutral current matrix (non-diagonal in
mass eigenstate neutrinos) \cite{Schechter:1980gr}.  In some models
these NSI may be sizable and lead to flavor and CP violation, even
with strictly degenerate massless neutrinos \cite{NSImodels2}.
On the other hand FC-NSI may also be induced by the exchange of
spinless bosons, a situation that arises in radiative models of
neutrino mass \cite{Zee:1980ai} as well as low-energy SUSY models with
broken R parity \cite{rpnu-spo1,rpnu-exp,rpnu-spo2}.  They also arise
in some supersymmetric unified models \cite{NSImodels3}, where they
may be calculable by the renormalization group evolution. Finally,
majoron models of neutrino mass lead to pseudoscalar NSI which may
induce invisible neutrino decays \cite{Schechter:1981cv,V}.

In addition to such renormalizable NSI, gauge theories of neutrino
mass may also lead to NSI of higher dimension $\geq 5$, such as
Majorana neutrino transition magnetic moments which lead to the
Spin-Flavour Precession phenomenon~\cite{Schechter:1981hw}.  As we
will see shortly, this may affect solar neutrino propagation in an
important way~\cite{RSFP}, providing two solutions to the solar
neutrino puzzle, one of which is resonant \cite{Miranda:2000bi} while
the other is non-resonant~\cite{Miranda:2001hv}.

\begin{figure}[htbp]
  \centering
\includegraphics[scale=.24]{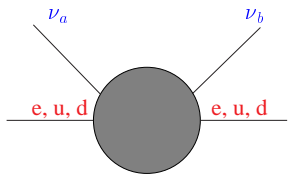}  
  \caption{Effective description of neutrino non-standard interactions.}
  \label{fig:nuNSI}
\end{figure}

In what follows I take an effective model-independent description of
NSI in which they are given as dimension-6 terms of the type
$\varepsilon G_F$, as illustrated in Fig. \ref{fig:nuNSI},
where $\varepsilon$ specifies their sub-weak strength. Such NSI may
affect the propagation of neutrinos in a variety of situations.
For example they have been considered both in context of the
solar~\cite{Valle:1987gv,Guzzo:1991cp,Bergmann:2000gp,Guzzo:2001mi}
and atmospheric ~\cite{Fornengo:2001pm,Gonzalez-Garcia:1998hj}
neutrino problems, as well as in connection with
astrophysics~\cite{Nunokawa:1996tg,Grasso:1998tt,Fogli:2002xj}.

In the light of such NSI one may ask, how robust are oscillations ?
To make a long story short, the answer is simple, atmospheric
oscillations are robust, solar are not. Let me start with the
atmospheric ones.

\section{Flavor Changing (FC) Neutrino-Matter Interactions}
\label{sec:flavor-changing-fc}

\subsection{Atmospheric \cite{Fornengo:2001pm}}
\label{sec:atmospheric}

Flavor-changing neutrino-matter interactions in the \nm-\nt channel
have been shown to account for the zenith-angle-dependent deficit of
atmospheric neutrinos observed in the SuperKamiokande experiment
contained events \cite{Gonzalez-Garcia:1998hj,Fornengo:1999zp},
without directly invoking neither neutrino mass, nor mixing.

\begin{figure}[htbp]
  \centering
\includegraphics[width=8cm,height=6cm]{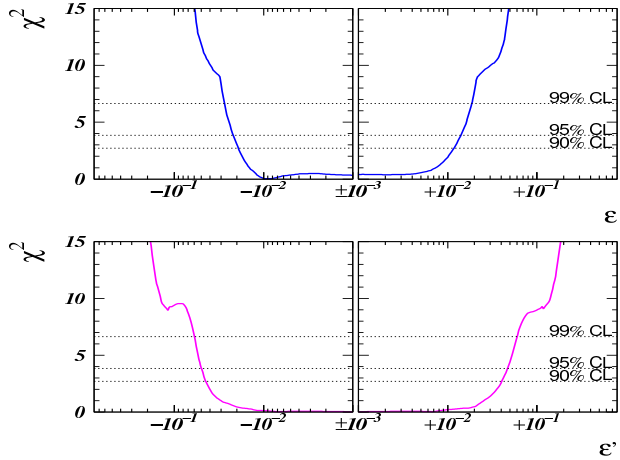}  
  \caption{Atmospheric bounds on non-standard neutrino 
    interactions, from \cite{Fornengo:2001pm}.}
  \label{fig:atmnsibds}
\end{figure}

However such FC explanation fails to reconcile the Super-Kamiokande
contained events with Super-Kamiokande and MACRO up-going muons, due
to the lack of energy dependence intrinsic of NSI conversions
\cite{Fogli:2000ak}.
The most recent global analysis of the atmospheric neutrino anomaly in
terms of NSI shows that a pure NSI mechanism is now ruled out at 99\%,
while the standard \nm $\to$ \nt oscillations provide a remarkably
good description of the anomaly \cite{Fornengo:2001pm}.
The robustness of the atmospheric \nm $\to$ \nt oscillation hypothesis
was shown to provide the most stringent and model-independent limits
on flavour-changing (FC) and non-universal (NU) neutrino interactions,
as illustrated in Fig. \ref{fig:atmnsibds}.
These limits are model independent as they are obtained from pure
neutrino-physics processes. The stability of the neutrino oscillation
solution to the atmospheric neutrino anomaly against the presence of
non-standard neutrino interactions illustrates the robustness of the
near-maximal atmospheric mixing and massive-neutrino hypothesis.

\subsection{Neutrino Factories-Atmospheric \cite{Huber:2001zw}}
\label{sec:nufact}

The proposed neutrino factories aim at probing the lepton mixing angle
$\theta_{13}$ with much better sensitivity \cite{Freund:2001ui} than
possible today by combining reactor and atmospheric data
\cite{Fornengo:2001pm}. Given the LMA solution of the solar neutrino
problem, it follows that there is a chance that indeed neutrino
factories (NUFACT, for short) will also probe the concurrent leptonic
CP violating effects (for example, through the measurement of CP
asymmetries).

\begin{figure}[htbp]
  \centering
\includegraphics[scale=.65]{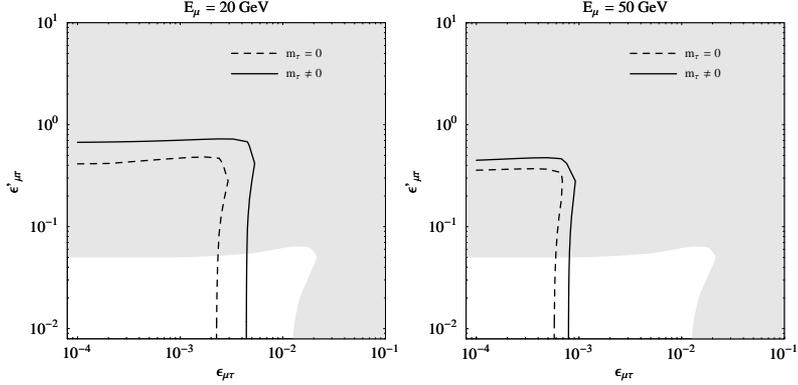}
  \caption{NUFACT sensitivities to non-standard neutrino 
    interactions, from \cite{Huber:2001zw}.}
  \label{fig:nufacnsibds}
\end{figure}

The potential of such a generic NUFACT in probing non-standard
neutrino-matter interactions has been determined in ref.
\cite{Huber:2001zw}.  It has been found that the sensitivity to
flavour-changing (FC) NSI can be substantially improved with respect
to present atmospheric neutrino data, especially at energies higher
than approximately 50 GeV, where the effect of the tau mass is small,
as illustrated in Fig. \ref{fig:nufacnsibds}.  For example, a 100 GeV
NUFACT can probe FC neutrino interactions at the level of few
$|\epsilon| < {\rm few} \times 10^{-4}$ at 99 \% C.L.

\subsection{Solar \cite{Guzzo:2001mi}}
\label{sec:solar}

Let us now turn to the solar neutrino oscillation hypothesis.  It
provides an excellent fit to the data and a powerful way to determine
neutrino mass and mixing. But, is it solid?  The lack of clear hints
of spectral distortion, day-night or seasonal variation in the solar
neutrino data implies that, despite the increasing weight of such
rate-independent observables, they do not give an unambiguous
smoking-gun signal for any particular solar neutrino conversion
mechanism.  Here I examine two viable alternatives which do not
require neutrinos to be mixed.

\begin{figure}[htbp]
  \centering
\includegraphics[width=3.6cm,height=3.6cm]{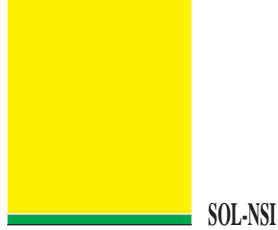}
  \caption{Hybrid three-neutrino interpretations of current neutrino data.
    The figure illustrates solutions to the solar and atmospheric
    neutrino anomalies in which there is only one oscillation, the
    atmospheric one, while the solar conversions are due to
    non-standard interactions (NSI).}
  \label{fig:hyb}
\end{figure}

I will show how one can have hybrid three-neutrino interpretations of
the current neutrino data in which the atmospheric data are accounted
for (mainly) by oscillations, while the solar neutrino data are
understood in terms of some sort of NSI, as illustrated in Fig.
\ref{fig:hyb}.

\begin{figure}[htbp]
  \centering
\includegraphics[width=8cm,height=7cm]{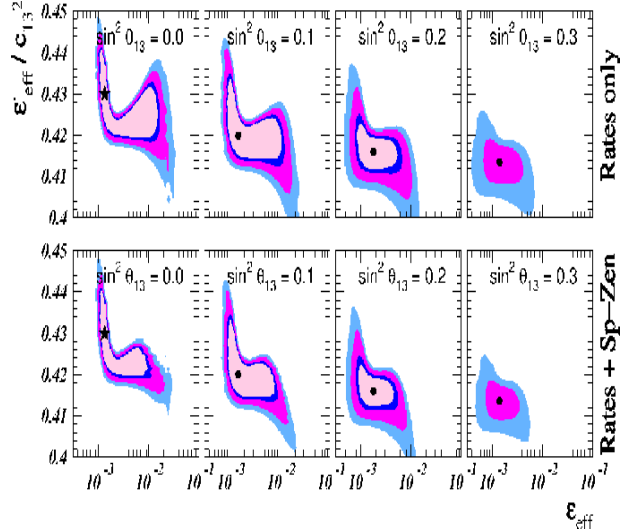}  
  \caption{Parameters of NSI solution to the solar neutrino anomaly, 
from \cite{Guzzo:2001mi}.}
  \label{fig:hybparam}
\end{figure}

FC-NSI solutions to the solar neutrino problem have been considered in
detail in ref.~\cite{Guzzo:2001mi}. 
Since the NSI solution is energy-independent the spectrum is
undistorted, the global analysis observables are the solar neutrino
rates in all experiments and the day-night measurements.
It was found that the NSI description of solar data is slightly better
than that of the oscillation solution and that the allowed NSI regions
are determined mainly by the rate analysis \cite{Guzzo:2001mi}.
By using simplified ans\"atze for the NSI interactions it was
explicitly demonstrated that the NSI values indicated by the solar
data analysis are fully acceptable also for the atmospheric data.
The required parameters are indicated in Fig. \ref{fig:hybparam}.

Since within such a three-neutrino scheme the solar and atmospheric
neutrino sectors are connected not only by the neutrino mixing angle
$\theta_{13}$ but also by the flavour-changing parameters accounting
for the solar data, it follows that to some extent NSI and
oscillations may be confused.

\subsection{Neutrino Factories-Solar \cite{Huber:2001de,Huber:2002bi}}
\label{sec:neutrino-factories}

The impact of NSI on the determination of neutrino mixing parameters
at a neutrino factory using the so-called ``golden channels'' \ne to
\nm and \bne to \bnm for the measurement of $\theta_{13}$ has been
given in ref. \cite{Huber:2001de,Huber:2002bi}.  It was shown how a
certain combination of FC interactions in neutrino source and earth
matter can give exactly the same signal as oscillations arising due to
$\theta_{13}$. This implies that information on $\theta_{13}$ can only
be obtained if bounds on NSI are available.  Taking into account the
existing bounds on FC interactions, this leads to a drastic loss in
sensitivity in $\theta_{13}$, of at least two orders of magnitude. A
near detector at a neutrino factory offers the possibility to obtain
stringent bounds on some NSI parameters. Such near-site detector
constitutes an essential ingredient of a neutrino factory and a
necessary step towards the determination of $\theta_{13}$ and
subsequent study of leptonic CP violation.

Additional signatures of theories leading to FC interactions would be
the existence of sizable flavour non-conservation effects, such as
$\mu \to e + \gamma$, $\mu-e$ conversion in nuclei, even in the
massless neutrino limit~\cite{Valle:1990pk}.

\section{Solar Spin-Flavour Neutrino Precession 
\cite{Miranda:2000bi,Miranda:2001hv}}
\label{sec:solar-spin-flavour}

\begin{figure}[htbp]
  \centering
\includegraphics[width=7cm,height=3.6cm]{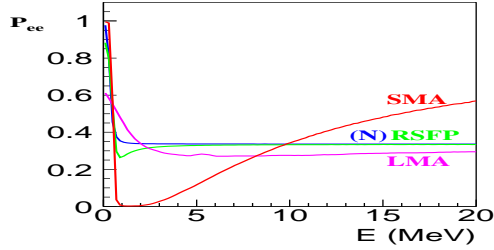}  
  \caption{Best fit oscillation vs spin 
    flavor precession \ne survival probability, from
    \cite{Miranda:2001hv}.}
  \label{fig:NMMprob}
\end{figure}

The Spin-Flavour Precession (SFP) of neutrinos \cite{Schechter:1981hw}
can be resonant in matter \cite{RSFP} and this leads to a solution of
the solar neutrino anomaly. The magnetic field profile may be fixed
self-consistently by physical requirements and by demanding it to
solve the solar magneto-hydrodynamics (MHD) equations
\cite{Miranda:2000bi}. The resulting \ne survival probability is
displayed in Fig.  \ref{fig:NMMprob} and compared with the one
expected in the case of neutrino oscillation.
There are two solutions: resonant (RSFP) and non-resonant (NRSFP).
Clearly both provide good global descriptions of the solar neutrino
data. 
\begin{figure}[htbp]
  \centering
\includegraphics[width=5.5cm,height=6cm]{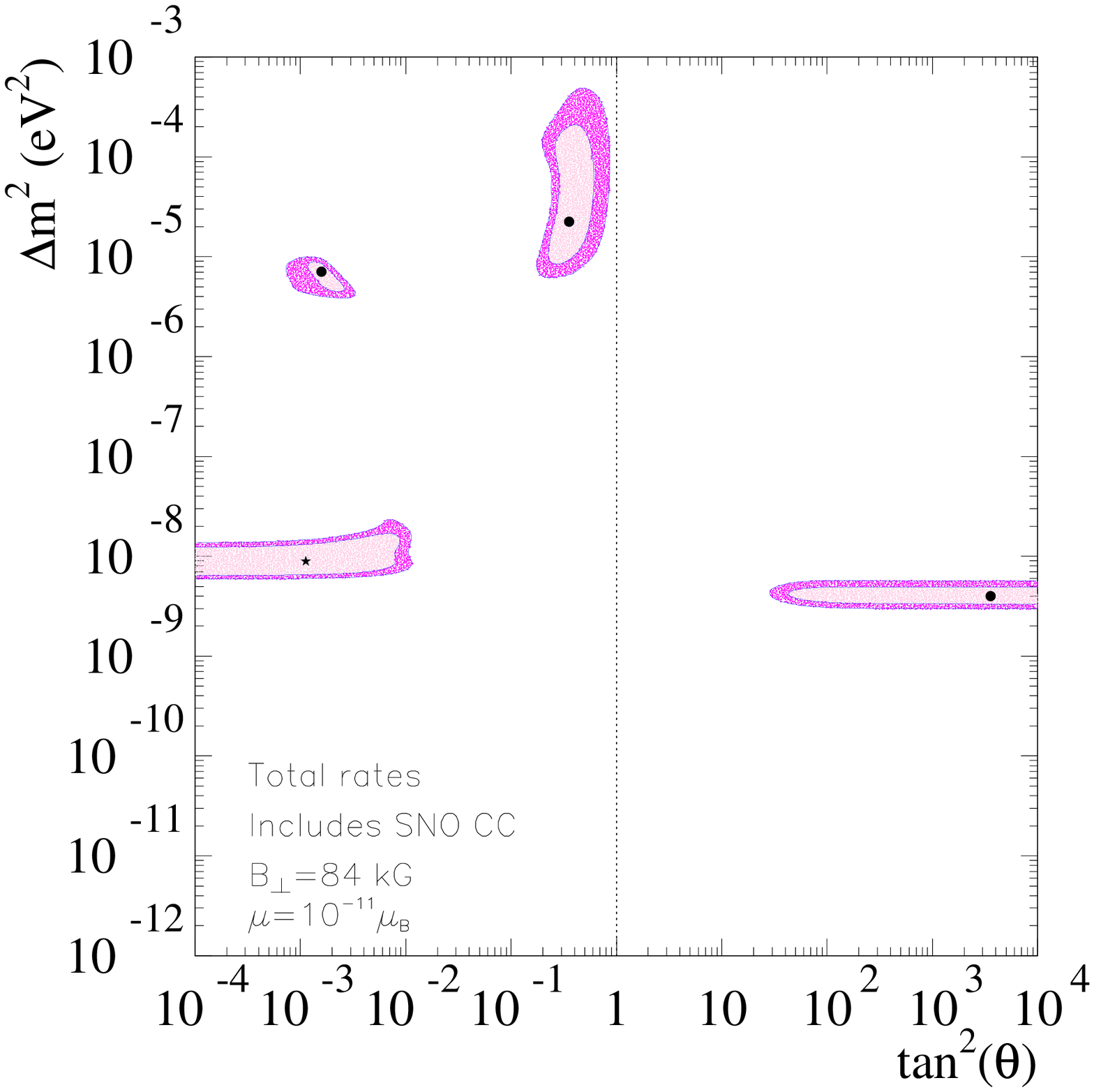}
\includegraphics[width=5.5cm,height=6cm]{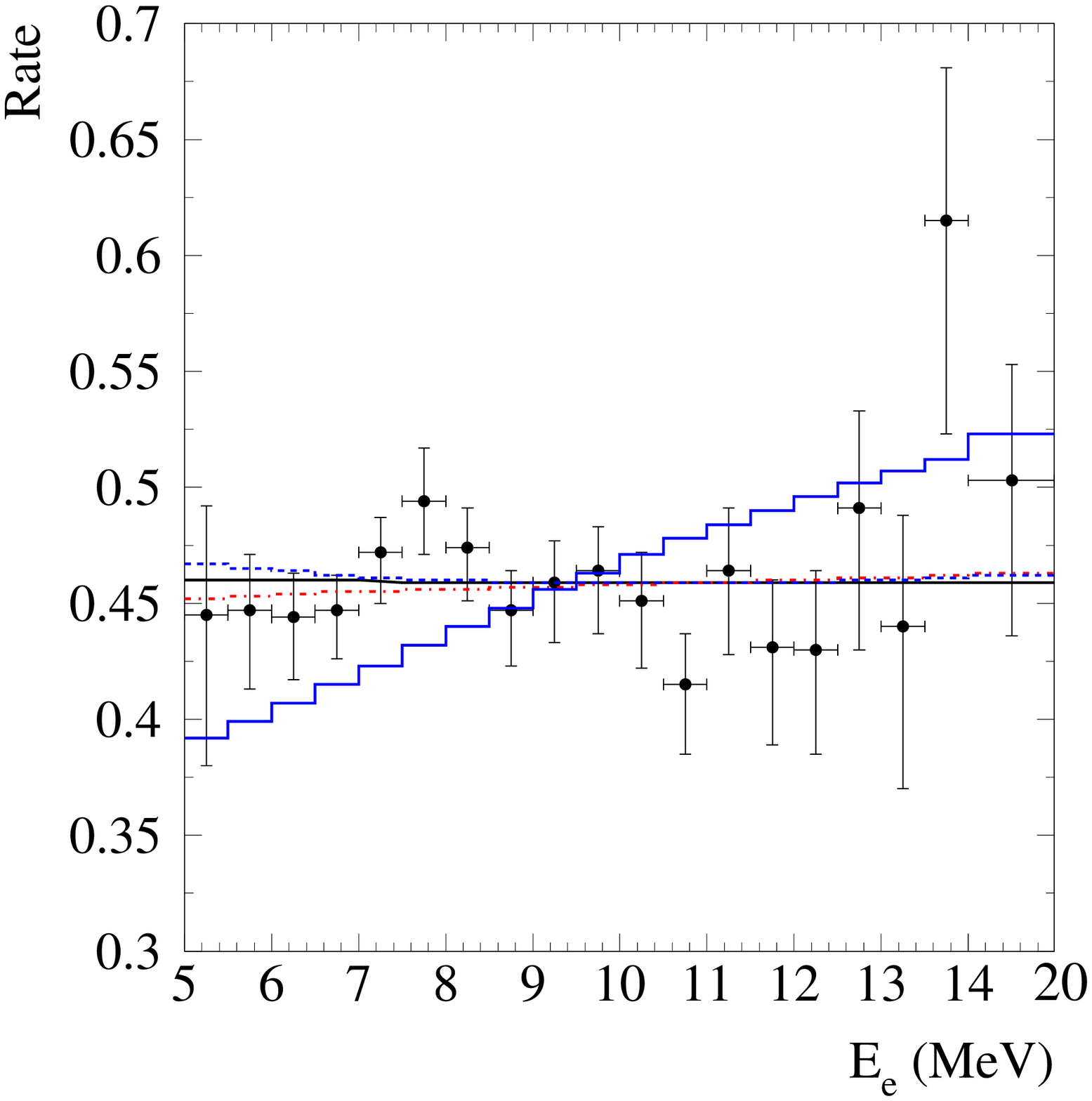}  
  \caption{``Grand Unified'' solar spin-flavor-precession/oscillation 
    plot and resulting recoil electron spectra, including the
    unacceptable SMA spectrum, from \cite{Miranda:2001hv}.}
  \label{fig:sfparam}
\end{figure}
In the left panel of Fig. \ref{fig:sfparam} the SFP parameters that
account for the solar neutrino problem are presented (the plot
includes the first SNO CC result, as well as the 1258--day
Super-Kamiokande data). In the MHD approach the resulting scheme has
only three effective parameters: $\Delta m^2$, $\mu B_\perp$ and the
neutrino mixing angle $\theta$.  The rates-only analysis for fixed
$\mu B_\perp$ slightly favors SFP solutions over oscillations. In
addition to the resonant solution (RSFP), there is a non-resonant
solution (NRSFP) in the ``dark-side''\cite{Miranda:2001hv}.  Note that
in the presence of a neutrino transition magnetic moment of $10^{-11}$
Bohr magneton, a magnetic field of 80 KGauss eliminates all large
mixing solutions other than LMA, as they would violate the upper limit
on anti-neutrinos from the Sun \cite{Aglietta:zu}.
The right panel in Fig. \ref{fig:sfparam} shows that the predicted
LMA, RSFP and NRSFP recoil energy spectra are in excellent agreement
with the data, in contrast to that of SMA.
Finally I note that it has recently been shown that the SFP solution
of the solar neutrino problem can be distinguished from the currently
favored oscillation solutions at Borexino \cite{Akhmedov:2002ti}.

Before closing let me also add that the possibility of fast neutrino
decays~\cite{V} has also been advocated could also play a sub-leading
role in the solar~\cite{Bahcall:1986gq} and/or atmospheric neutrino
problems~\cite{atmdecay}. The data allow for neutrino stability tests,
though not specially stringent.

\section{In a Nutshell}

\begin{figure}[htbp]
  \centering
\includegraphics[width=10cm,height=6.5cm]{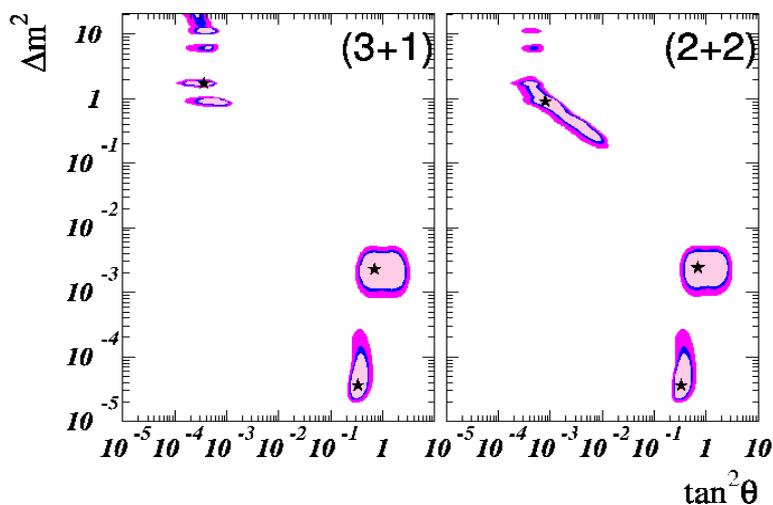}  
  \caption{``Grand-unified'' neutrino oscillation plot, adapted 
    from ref.  \cite{Maltoni:2001bc}, before
    \cite{Ahmad:2002jz,Ahmad:2002ka}.}
  \label{fig:4nuoscplot}
\vglue -.3cm
\end{figure}

It is now very hard to dispute the need for physics beyond the
Standard Model in order to explain current solar and atmospheric data.
In order to also account for the LSND hint \cite{lsnd} indicating \bnm
to \bne oscillations with a eV-mass-squared difference in the
framework of quantum field theory one needs an additional light
sterile neutrino \ns \cite{giunti,Caldwell:kn,ptv92,pv93}.
Currently LSND is neither confirmed nor ruled out by any other
experiment.
A unified global analysis of current neutrino oscillation data for
4-neutrino mass schemes has been given in \cite{Maltoni:2001bc} using
data from solar and atmospheric neutrino experiments, as well as
information from short-baseline experiments including LSND. 
The summary of current 4-neutrino oscillation parameters is given in
Fig. \ref{fig:4nuoscplot}.
One finds that the solar and atmospheric data reject sterile
neutrinos.  For example, recent SNO data rule out conversion to
\texttt{ pure} sterile neutrino states at the $\sim 5\sigma$ level. A
smaller degree of sterile neutrino rejection is found for atmospheric
data \cite{Maltoni:2001bc}.  For this reason, the best 2+2 model model
has now changed from one with the sterile state mainly in the solar
sector \cite{pv93} to one in which it lies mainly in the atmospheric
sector \cite{ptv92}.  However I stress that, in a truly global fit
including also the LSND data \cite{Maltoni:2001bc}, sterile neutrinos
can not be ruled out, since the latter disfavor the Standard Model at
99.9\% CL with respect to the best 4-neutrino model. The best-fit
four-neutrino model is a 3+1 scheme, since in this case one can
decouple \ns from both solar and atmospheric sectors.  However, in
such a global sense, even 2+2 models of the type considered in ref.
\cite{ptv92} can not yet be ruled out, being still allowed at the 90\%
CL \cite{solat02}.  For the moment the sterile neutrino oscillation
hypothesis is therefore alive, though there is a clear tension between
underground and short-baseline data.  We wait anxiously for results
from MiniBooNE to settle the issue~\cite{MiniBooNE} in the near
future.

\par

I thank Martin Hirsch for comments on the manuscript.  This work was
supported by the European Commission grants HPRN-CT-2000-00148 and
HPMT-2000-00124, by the ESF \emph{Neutrino Astrophysics Network} and
by Spanish MCyT grant PB98-0693.

\end{document}